# PyDraw: a GUI drawing generator based on Tkinter and its design concept

Jinwei Lin[1]    Aimin Zhou[2]

**Abstract** The emergence of GUI is a great progress in the history of computer science and software design. GUI makes human computer interaction more simple and interesting. Python, as a popular programming language in recent years, has not been realized in GUI design. Tkinter has the advantage of native support for Python, but there are too few visual GUI generators supporting Tkinter. This article presents a GUI generator based on Tkinter framework, PyDraw. The design principle of PyDraw and the powerful design concept behind it are introduced in detail. With PyDraw's GUI design philosophy, it can easily design a visual GUI rendering generator for any GUI framework with canvas functionality or programming language with screen display control. This article is committed to conveying PyDraw's GUI free design concept. Through experiments, we have proved the practicability and efficiency of PyDrawd. In order to better convey the design concept of PyDraw, let more enthusiasts join PyDraw update and evolution, we have the source code of PyDraw. At the end of the article, we summarize our experience and express our vision for future GUI design. We believe that the future GUI will play an important role in graphical software programming, the future of less code or even no code programming software design methods must become a focus and hot, free, like drawing GUI will be worth pursuing.

**Jinwei Lin**

linjinweiyouxiang@163.com

**Aimin Zhou**

amzhou@cs.ecnu.edu.cn

[1] Electronic science and technology major, School of physics and materials science, East China Normal University, Shanghai, China

[2] Department of Computer Science and Technology, East China Normal University, Shanghai, China

# 1   Introduction

Since the onset of the screen, the graphical user interface (GUI) has been an important means of human-computer interaction between computer black boxes and the outside world. Through GUI, users can easily communicate with the computer and transmit task instructions to the computer. At the same time, the computer can also visually display the task results to the user through the GUI. GUI is an important part of human-machine interaction software.

Python is a widely used programming language, in the IEEE release of 2017 programming language rankings: Python ranks first(Stephen 2017), and in 2018 Python still ranks No. 1(Stephen 2018). According to TIOBE Index for August 2018 (TIOBE 2018), Python has a strong upward trend and is now ranked fourth. Python has many excellent programming features and charm, but also attracted a large number of people to become its users(Avnish 2018). Current Python has a large number of practical and efficient libraries(ThePSF 2018), performing very good in scientific computing and data processing, especially in the current hot AI field. However, users of Python often encounter the problem that Python's GUI design is a weakness of Python. As an elegant programming language, Python should also be able to design a GUI elegantly to communicate with users. The good news is that Python itself has its own framework as GUI design libraries, Tkinter, which allows any environment that can run Python to write GUIs through Tkinter and present the same native UI interface. But users who use Tkinter will soon find that Tkinter is not visualized, although it has multiple layout cubes, However, the simple direct coding method can not achieve real-time visual GUI design. And we know that the users using Tkinter don't really want to design a large 3D game, more often than not, they want to design a simple and convenient GUI application. As Python's de facto standard GUI library, the Tkinter GUI framework will be the first choice for designing easy and concise GUIs.

Python is a charming language and Tinter is a charming GUI framework. As a loyal fan of tkinter, CameronLaird (wiki.python.org) also calls the yearly decision to keep TkInter as "one of the minor traditions of the Python world." It can be seen that the location of Tkinter's chief GUI library has been confirmed long ago. So, when the user decide to choose Tkinter as his preferred GUI design framework, and roll up the sleeves and get ready to get some big things done, the reality will remind him that he need to keep typing instructions code with his keyboard unless he have an automatic GUI generator that can greatly speed up his work, as shown in Table 1.

In order to have a clear understanding of PyDraw, we will also list PyDraw above as a comparison. According to our survey, the GUI generators with more users, mainly

**Table 1** The main Python automatic GUI generators on the market and comparison with PyDraw.

| Design Tool | Latest Version | Latest Update | Basis | Other Packages | Package Size | Note |
|---|---|---|---|---|---|---|
| Komodo | 11.1.0 | Uncertain | Uncertain | No | 90.02MB(mis) | Commercial, Multi platform |
| PAGE | 4.14 | 2018/5/31 | tkinter/Tcl | Yes | 9.47MB(zip) | Free, Full platform |
| pygubu | 0.9.8.2 | 2018/1/23 | tkinter | No | 1.33MB(zip) | Free, Full platform, Code Add |
| PythonWorks | 1.3 | 2002/11/25 | tkinter | Uncertain | Uncertain | No longer being developed |
| xRope | 1.44 | 2013/4/17 | tkinter | No | 2.07MB(zip) | Free, Full platform |
| Visual Tkinter IDE | 2v0.12 | 2013/4/26 | VB,tkinter | Yes | 5.21MB(zip) | Free, Windows |
| PyDraw | 1.0 | 43313 | tkinter | No | 295KB(normal) | Free, Full platform |

Komodo, PAGE, Visual Tkinter IDE and pygubu. Regrettably.
(https://wiki.python.org/moin/GuiProgramming)

    PythonWorks has stopped its technical support and update. The most powerful Komodo is a commercial version. PAGE needs to install additional Tcl libraries when it is used. The performance of pygubu is similar to that of PyDraw, but it is not simple enough to use. xRope is good to use, but it follows the (GPLv2) protocol. Visual Tkinter IDE is designed on the VB platform. It has to be implemented with VB. So users are limited to Windows users. Of course, the above software toolsa all are excellent. PyDraw is designed to provide a simpler and more efficient GUI generator, and this paper is written for how to achieve this goal. PyDraw is still only V1.0 version, is the original version of the principle implementation, now only has the basic GUI generation function, although only V1.0 version, it has been able to complete the Tkinter all commonly used GUI control editing design and property binding. Most importantly, please look at Table 1. Its size is only 295KB. In fact, the V1.0 version of PyDraw has only one .py file, and we can say that in volume, the V1.0 version of PyDraw well inherits the simplicity and ease of use of Tkinter. In terms of GUI design performance, PyDraw also has excellent performance. Please refer to the introduction later. The focus of this paper, however, is not just on what PyDraw can do, but why PyDraw can do it, that is, how PyDraw works, and how to design a similar GUI generator for any programming language or framework that has Grath or Canvas capabilities. And this is, the design concept of UI generator in PyDraw. Next, this paper will show the design thinking and process of PyDraw, as well as the design concepts behind it, which has greater value.

    After reading this paper, we hope the readers have the following abilities and achievements:

1. Know that there is a GUI design tool called PyDraw.
2. Know how to use PyDraw to complete a succinct and elegant GUI design.
3. Know the design and operation principle of PyDraw, and the construction concept of GUI design tools embodied by PyDraw.
4. Know how to use PyDraw's design philosophy to upgrade any programming language or framework which has Canvas parts to a GUI generator tool level.

The purpose of this paper is not only to clearly show the design principles and ideas of a Python standard GUI library design tool, but also to express the design concepts behind it. Scientific design concepts are the most valuable, and also the core idea of this paper to convey.

## 2  Background and related work

In this section, we will elaborate the story behind PyDraw, and the design principle of PyDraw and its functional implementation. And what is the design concept behind PyDraw.

### 2.1  The meaning behind the name of PyDraw

At the beginning of writing PyDraw, many people were asking why the tool called PyDraw was not called by any other name, such as PyGUI, and so on. Indeed, if look at the functionality and implementation of PyDraw alone, this tool should be called PyGUI. Because PyDraw is used to generate GUI, it is a GUI tool software. But we ended up choosing PyDraw because we felt that simply building a GUI for users was clearly not the ultimate goal of GUI design. We named PyDraw to express a simple and convenient GUI tool, but also to express the idea of freely drawing beautiful and practical GUIs, like drawing. By simple click and drag, through simple lines and images, easy to create an ideal GUI, without writing any code, automatically generate GUI-related software, or even non-GUI utility software, is the highest level of GUI design. In this level of GUI design, GUI has become a programming tool, not just a way to achieve human-computer interaction interface. PyDraw pursuit is free to draw, this is a pursuit of the beauty of design art. We want to express the concept of "all GUIs are by drawn" through PyDraw. "By Drawn" is the core of "PyDraw". Although PyDraw's current GUI design is not beautiful enough, in the future, more advanced versions of PyDraw will bring artistic beauty to the user's design experience.

## 2.2 PyDraw main construction principle and control realization

PyDraw exists as a lightweight Tkinter automatic GUI generator. The core design principle of PyDraw is that using the control and graphics drawing function of Canvas Widget in Tkinter framework as the whole Canvas. The "Canvas" is also the GUI design emulational window, that is to say, The whole Canvas is a simulated GUI window, and the whole Canvas will be used as a container to host other controls drawn on it. We called this drawing board GUI Design Canvas. What the GUI users see in GUI Design Canvas is the real GUI generated by GUI Code after transformation. As shown in Fig. 1, Fig. 1 has a name called the "Eyes of Pydraw", which represents the design idea of seeing the wonderful world builed by Tkinter controls through PyDraw implementation. Obviously, in PyDraw's architecture, Tkinter's Canvas controls are indispensable. Through research and analysis, we can find that PyDraw is actually a Tkinter automatic GUI generator implemented by Canvas of Tkinter. That is to say, the operating interface of PyDraw is mainly written by Tkinter, the visual control drawing process is also implemented by Tkinter, and the transformation and generation of GUI Code is also based on Tkinter. So PyDraw is a highly native GUI design tool that supports tkiniter, and that's one of the reasons why the size of PyDraw just more than 200 KB. Because it is native support, there are not third party libraries that must be installed. To a large extent, this can saves volume. The GUI design process of PyDraw has the characteristics of WYSIWYG, and is native to support Tkinter. Because Tkinter's Canvas doesn't have the ability to draw windows, we use the entire Canvas directly to simulate the GUI window, and add a title bar and a title button to make the GUI Design Canvas look more like the resulting GUI window. This method is also more in line with PyDraw's design concept, "all GUIs are by drawn".

The Tkinter framework is great, but users familiar with Tkinter may have an experience that Tkinter provides too few basic controls. Sometimes it can not meet the actual design requirements. So, on the control set, PyDraw introduces a TTK control support, that is the "Combobox," a drop-down list, which is a commonly used control. Other TTK controls can also be introduced by the the same way. As shown in Fig. 1, PyDraw currently supports a set of controls and all of them are most commonly used by Tkinters, including Button, Canvas, Combobox, Checkbutton, Entry, Frame, Label, Label Frame, Listbox, tk. Message, PanedWindow, Radiobutton, Scale, Spinbox, Text, and Menu. Among them, we need to pay special attention to the "Menu" control, Menu control can not be directly drawn on GUI Design Canvas, in order to achieve the purpose of drawing Menu control, we designed a PyDraw Menu editing control, specifically for editing the properties of the Menu control in the GUI. The usual control "Scrollbar" is not added to PyDraw, because in Tkinter, Scrollbar is added to some one control directly, and there must be a control be specified as a carrier for Scrollbar, such as Listbox, Canvas, or Text, as well as windows. The additions of Scrollbar should not be achieved through dynamic painting, but should be added directly by clicking. For such controls, the user can add them by selecting the carrier to add. Scrollball will be added in the later version of PyDraw.

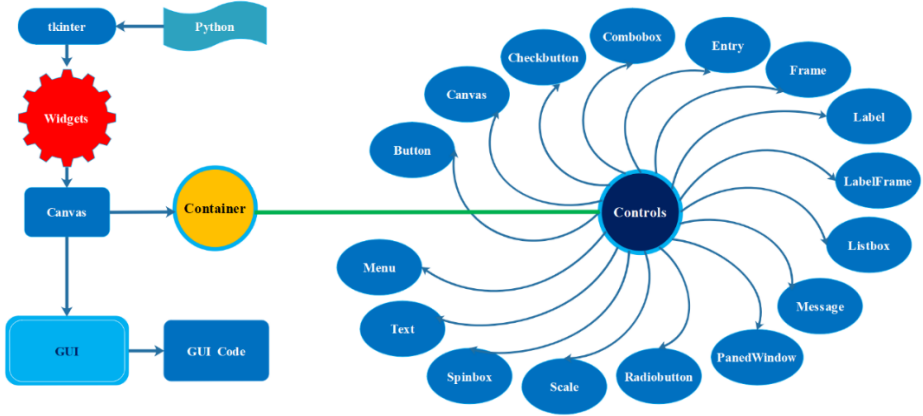

**Fig.1** The main construction principle and control realization of PyDraw

In the blueprint of PyDraw, GUI Design Canvas plays an important cornerstone role. It is the foundation of the whole GUI design. Because of the property setting of Canvas on Tkinter, the area between any two controls on Canvas can not overlap. This makes the user more objective and rational when drawing GUI. However, this also makes it difficult to lay out the control by clicking and dragging on any control that can act as a carrier on Canvas as the parent container. On the other hand, this also reflects the shortcomings of PyDraw, so we can see that if all the future development of PyDraw is based on Tkinter, then the development of PyDraw will be greatly constrained by the development of Tkinter. Fortunately, we have not only studied the principles and steps of GUI designing by PyDraw, but also explored the design ideas behind it. PyDraw's design ideas will guide us to easily add other GUI design framework to PyDraw, that is, PyDraw can evolve.

Later versions of PyDraw need not be restricted to the Tkinter framework. We say PyDraw is an evolutionary GUI design software because PyDraw's design philosophy is universal and effective. In the following chapters, we will explain this excellent GUI design concept in detail.

## 2.3  The design concept behind PyDraw

This is not only the discuss core of this paper, but also the core concept of PyDraw design, which is a general design concept. If this idea is expressed in one sentence, it can be described as " all GUIs are by drawn, and Canvas is the key". Indeed, if the user want to paint something, Canvas is very important. Although this sentence is concise, it contains a lot of information. "By Drawn" is the key of PyDrawn.

As we know, the screen is an important tool for displaying data and information, and also an important way for users to interact with computers. In the early computer, users could interact with the computer by inputting instruction codes, but the efficiency of this interaction is very low, and the general instructions are difficult to remember. The emergence of GUI has completely changed this situation and improved the efficiency of human-machine interaction. Experience tells us that graphic information is more intuitive than characters information. The emergence of GUI has also greatly promoted the development of computers. In modern times, GUI is already an important way to achieve communication between human beings and computers.

To understand the first half of the core idea behind PyDraw, "all GUIs are drawn." From the most primitive GUI generation point of view, suppose the screen display is just a board with many pixels of light point, the black-and-white display can only display grayscale colors between black and white. Color displays can display colorful colors. If the computer doesn't control the display of the board, it's probably just a white light-emitting panel, or some other color, or even a direct black screen. When the screen is a whiteboard, there is no information, and the user can not communicate with the computer through the screen. So to communicate with the user, the computer must let this screen display some information that the user can recognize, such as graphics or text. When a computer can interact with a user through certain tags such as images or text, the GUI starts to function. So, in the face of "all GUIs are drawn" it's easy to imagine how the GUI was drawn, and who drew it?

In the broad sense, the painters here has two, one is the computer, the other is the user. The first type of painter is the one that no direct intervention from human beings. When a computer draws a GUI, it sends instructions to the screen according to a pre-programmed program or non-user direct instructions it receives, and changes the display state of some pixels on the screen. If the pixels that receive computer instructions on the screen just gather into a square, a square appears on the screen, which is the signal that the computer sends through the screen. Signal is just the message, it is the medium of communication between people and computer. Whether it is a simple period character or a dynamic image that requires a powerful GPU to interpret and process, it is inseparable from the precise control of the screen pixel set by the computer. The second type of painter is the GUI painter rendering of human direct intervention. Man sends instructions to the computer. The computer draws GUI according to the contents of user instructions. That is to say, the computer is the direct drawer and the user is the indirect drawer. So no matter what kind of rendering, the computer is the direct executor of rendering the GUI, the key is whether the user is involved. That is, as long as the user know the driver instructions for the computer to draw the GUI, then everything in the world can be displayed through the users' instructions. From a small to almost invisible period character, to a bright and boundless cosmic star.

Using a computer to draw a GUI is to select a number of points on the screen, make they aggregate to display different colors to distinguish from other points, and these points aggregate areas can express the information a user can understand. Then, these special display areas have specific event response. By typing in hardware, such as keyboard, mouse, screen touchpad and so on, the user can focus on the display area triggered by the user's touch or click, that is, focusing and getting the focus point. While the selected area changes in display, distinguish it from other areas not selected by the user. And these areas, that is, the control area, when the cursor enters the control area again, if triggered at the same time, will start the corresponding trigger event. That is, the computer drawing GUI has two main parts, the first is a specific graph, the second is a specific trigger event.

So how do users draw GUI through computers? Generally speaking, GUI is just windows and controls. To draw a window, it is to display a fixed window style on the screen. The type of window in Tkinter is set, and there's not much change, so the GUI rendering feature for the primary version of PyDraw focuses on the control's rendering. "All GUI is drawn, and Canvas is the key". Generally speaking, the canvas of a window is the screen and the canvas of the controls drawn on the window is the window. That is to say, canvas is the carrier of GUI components. In the computer world, a window is canvas for the controls on the window, while the screen can display only the user's visible area relative to the entire screen. In visual situations, the screen is the largest Canvas, and after the invisible part is included, the area behind the screen is unlimited, and the user can see only a small part of it. Canvas can show everything. So we say that all GUI is drawn and canvas is the key.

Next, we'll illustrate the implementation of PyDraw's underlying principle, which is the GUI rendering process behind the PyDraw concept, in conjunction with Fig. 2. First, the user should choose an area on the screen as a window and make it look like a window to distinguish it from other areas. The implementation of this step needs to invoke the control function of the system to screen display. After determining the display area of the window, enter the "click and drag process", which is the process of drawing the control. For example, in the blue rectangular box in Fig. 2, the user can control ther input hardware, select a point on the screen, and we will select the mouse to explain.

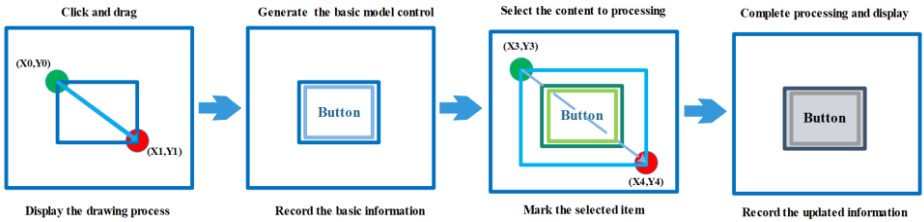

**Fig. 2** The process of drawing GUI on screen by PyDraw

*Phase one:* click drag and drop. The mouse clicks on a point in the window as a *P0 (X0, Y0)* point. In PyDraw, the *P0 (X0, Y0)* point is the upper right corner of the control, and it is also the positioning point when the control carries out the layout of ".*place()*". After clicking the left mouse button, hold down the left mouse button, the cursor will become a cross marking state, drag the mouse, move the cursor, release of the left mouse button after reach *P1 (X1, Y1)* point, complete the choice of the space to draw in the real area. The *P1 (X1, Y1)* here is the bottom right coordinate coordinate of the control to be drawn. Then the system will record the coordinate data of the two points of *P0 (X0, Y0)* and *P1 (X1, Y1)*. It should be noted that the coordinates here, as well as the coordinates involved in the next third phase, are all originated from the upper left corner of GUI Design Canvas *(0, 0)*. The specific coordinate settings are shown in Fig. 3. The X and Y directions are right and down, respectively. Why do we need to record these two coordinate data, because there are the following relations:

$$width = abs(X1 - X0)$$
$$height = abs(Y1 - Y0)$$

Because width and height are positive numbers, we use "*abs()*" to get their absolute value. That is, with a simple click-and-drag operation, we can determine the width and height of the control to be drawn and the position of its coordinates. The process of drawing a GUI on a computer is so efficient and elegant that PyDraw's idea is that when the user can easily draw a GUI, the processing mechanism will silently saves the precious data the user need in the background. In PyDraw, the whole process of drawing the control is dynamic, that is, the user can observe the changes of the control every moment in real time when the user draw the control on UI Design Canvas. This also shows that in the process of drawing a GUI, the shape of the control being drawn will be updated instantly with each slight drag of the user. That is to say, PyDraw's GUI control drawing is real-time, dynamic and visual. At each time the user clicks and drags, the shape of the control will be updated regularly and instantly.

To achieve this step, the real-time visual updates designed in PyDraw are as follows: First, click a point on UI Design Canvas to create a preliminary control prototype. If the user does not drag it after clicking, the prototype of the control is generated on UI Design Canvas. If the user holds down the left button after the left mouse button is clicked and dragged, the internal processing mechanism of the program will trigger a configuration event in each cursor displacement operation caused by the drag to update the width and height of the drawn control. The entire rendering process will be displayed on UI Design Canvas.

**Phase two:** generate basic model controls. In operation, the user holds down the left mouse button and drag, to get their own ideal drawing control, then release the left mouse button, at this time the UI Design Canvas will generate the basic model control base on the user's preliminary idea. Processing mechanism of the program will start the data storage program at this time to store the data information of the basic model control. The information storage task of the drawing control is triggered only after the drawing of the basic model control is completed. The purpose of this design is to avoid the waste of computer computing resources. After releasing the left mouse button, the data information is needed by the program, so the data storage task should be done after releasing the mouse.

**Phase three:** Select the control to processing. In the last phase, the user has completed the drawing of the underlying model control. If the user is not satisfied with other properties of the drawn control, then the user can modify the properties of the control. However, the computer does not know that the user is or not satisfied with the control unless the user tells it. That is, the user needs to select a control to processing before modifying its properties, and this select and handle process is the phase three. In this phase, the user has to draw a rectangular circle around the selected control, analogous to the first stage, the left mouse click on the first point marked as (X3, Y3), press and drag to form a selection box, the left mouse button release point corresponding to the position of the canvas coordinates (X4, Y4), to determine the selected area .The processing mechanism will compares all the drawn controls one by one to find the controls contained in that area, and mark them to distinguish it from other controls.

**Phase four:** Finish processing and display. In this stage, the user changes the properties of the selected control, the program processing mechanism performs corresponding changes, and saves the updated control information, while updating and displaying in real time on GUI Design Canvas.

From the above discussion we can know that the process of designing a computer to draw a GUI just is the process of making full use of coordinate and graphic transformation. The use of coordinates is the use of pixels, the use of graphic transformation, is the use of event response.

"All GUIs are drawn, and canvas is the key." That is, the user can draw a GUI as long as the user find a way to make the pixel display of the corresponding coordinates change as specified, and in the process, if the user find the canvas widget the user can use, the user can write an automatic GUI generator. Canvas will make everything easy. Essentially, controls are just a special display of graphics, such as a button, made up of a bunch of pixel sets that come together to show the shape of the button. A group of pixels with a specific display, in the area formed by the edge of the button shape, with different display configuration and position ratio, show the shape of the button, thus obtaining a basic button display.

But it's not a complete button yet, because nothing will happens when the user click on it with the mouse, which means that the computer draws an image that looks like a button. In order for this image to function as our control, we must also define its trigger events. Similar to the real scene, we need to define the morphological transformation of the button that accepts a mouse click, and make the button have other attributes that we want to give it.Finally, we must give the function of connection with trigger event function to the button. For example, we want to define a single click event, from the computer program processing mechanism, is to define, when the mouse moves, the cursor also moves, and when the corresponding coordinates of the cursor into the control coordinate area, start the monitoring event, monitor whether the mouse is pressed at this time, and define the trigger when the cursor is pressed. That is, the mouse click event. At the same time, we have to define the pixel display of the corresponding area of the clicked control to display the effect of the click.

From the above analysis we know that, as long as there is a control method to control pixels displayed by coordinates, we can design the GUI. And canvas is the most important component and implementation part. With canvas, we can naturally update a GUI design framework to a visual GUI generator. PyDraw is born by this design concept. PyDraw wants to convey not only that it is a GUI generator based on the Python Tkinter framework, but also that this GUI design concept of human-computer interaction.

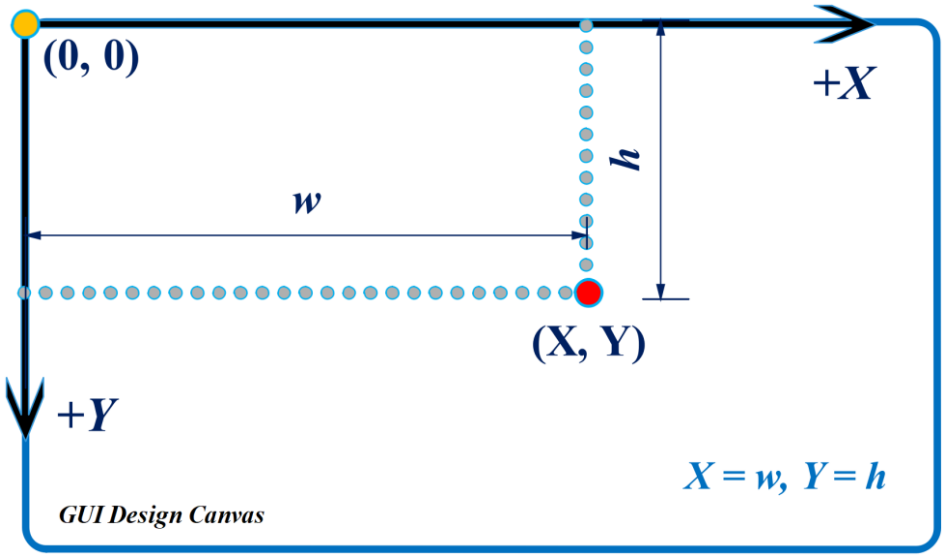

**Fig. 3**  The coordinate world on GUI Design Canvas

## 3 Use PyDraw to draw GUI

### 3.1 main interface

When the user click PyDraw, the user will enter the main interface of PyDraw. Fig. 4 shows the full development panels of the PyDraw design interface.

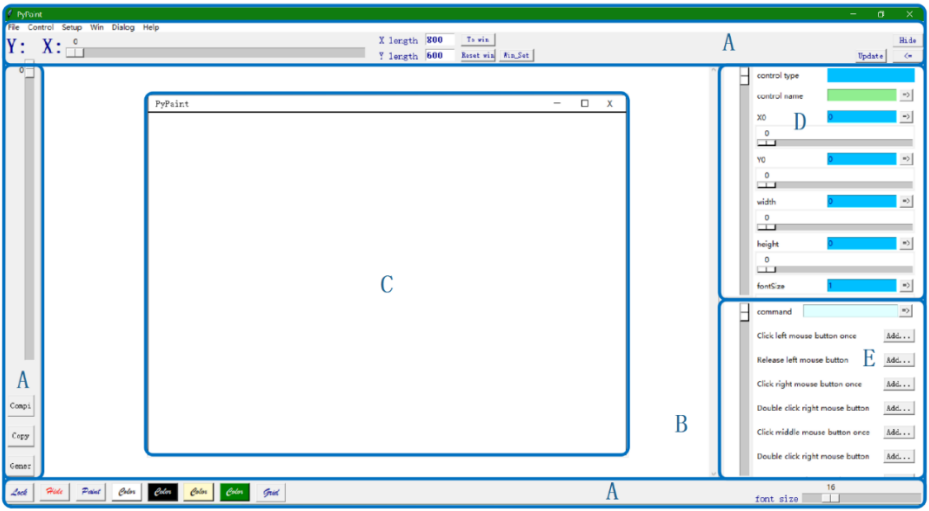

**Fig. 4**  Main page of PyDraw running with full development panels

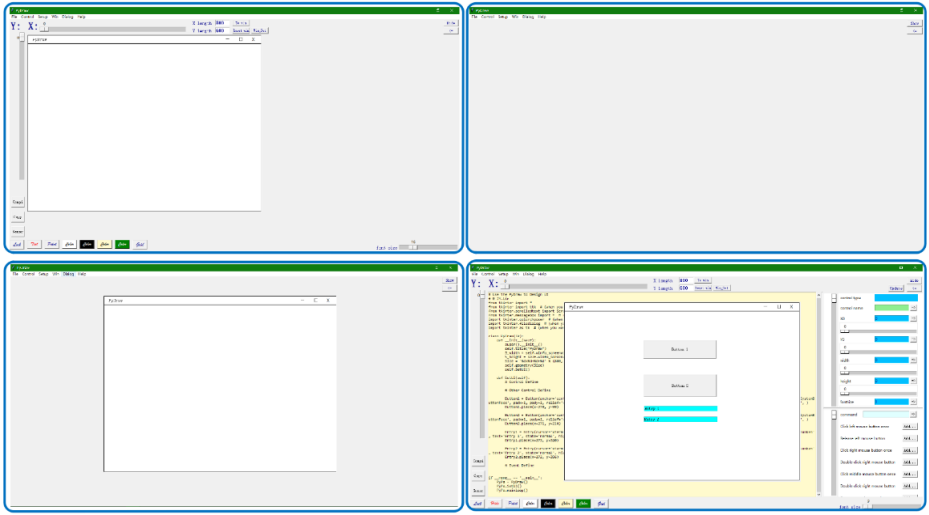

P1.the initial interface  P2.interface with all functional pages hidden

P3.only the GUI Design Canvas  P4.all the functional panels are displayed

**Fig. 5**  PyDraw main interface state display

There are 5 parts:

A: Control Panel, B: Converter Text Box, C: GUI Design Canvas, D: Properties Setting, E: Events Setting. The vast majority of PyDraw GUIs are designed with Tkinter native controls, and only one control, the drop-down list of uses a class "Combobox" of TTK controls. The GUI Design Canvas is designed to allow users to draw GUIs more freely and happily to design GUI, dragging GUI Design Canvas by pressing the middle mouse button. Properties Setting and Events Setting are both designed with Tkinter's Scrollbar and PaneWindow controls. PyDraw has three other main parts, Processing and Global Variables, and Window Setting.

Each of these five parts can play an important role. Control Panel: is the control panel, used to set the properties of PyDraw, which can assist the drawing of the GUI, set the width and height of GUI Design Canvas, setting Converter Text Box and display font, start GUI code conversion, set auxiliary grid, reset window and implement other functions. Converter Text Box is used to display converted GUI code. GUI Design Canvas: simulation window and GUI drawing area. Propeties Setting and Events Setting are propeties and events settings regions respectively.

Note that when start PyDraw, the B, D, and E areas are hidden, as shown in Fig. 5 P1. The ability to hide panels and control components is a feature of PyDraw. Each of A, B, C, D, E five main areas all can be hidden with clicking on the corresponding keys. The purpose of this design is to facilitate users in drawing the GUI, so that users can be less affected by PyDraw interface layout, with more space and a more pure mentality to achieve the design of the GUI. When need to design in a clean environment, use the scene shown in P3 of Fig. 5. When need to do a efficient design wiht a fully functional environment, use the scene shown in P4 of Fig. 5. C, D, E three areas have click-and-drag characteristics, D and E areas have sliding characteristics, which is to bring users a better using experience.

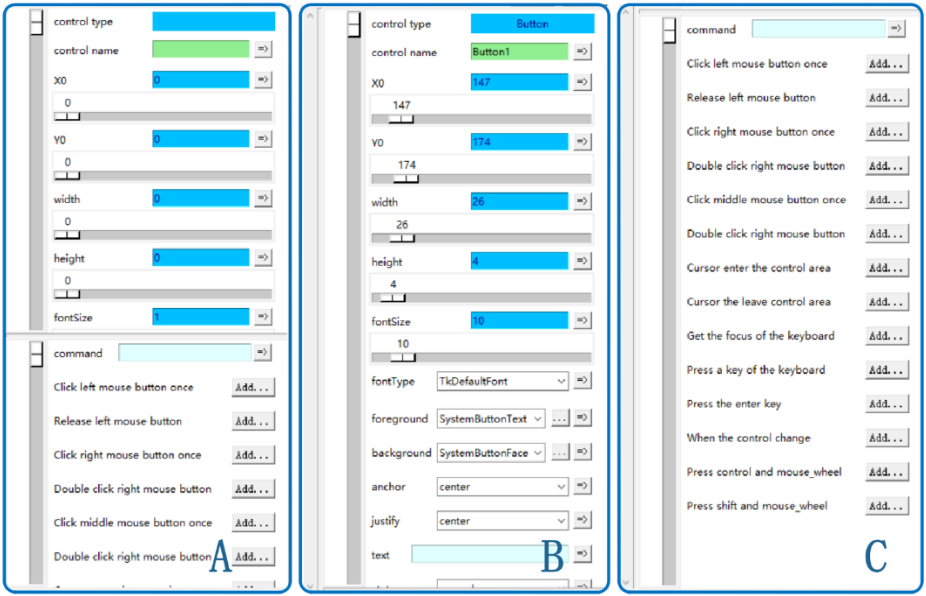

**Fig. 6** Attribute Setting and Events Setting composite panel

As shown in Fig. 6, when the D, E composite panel is opened, the A subgraph is displayed, the B subgraph is dragged down the boundaries to the end, so that the D region has the largest visual area, the C subgraph is dragged up the boundaries to the end, so that the E region has the largest visual area. In addition to changing the visual area of the D and E plates by dragging the boundaries, the user can also change the visual area by moving the cursor on the left scale control and rolling the mouse roller up and down. We can see that the basic properties and events have been added to PyDraw.

## 3.2　Drawing steps

It is very convenient and quick to draw GUI through PyDraw. The main steps are:

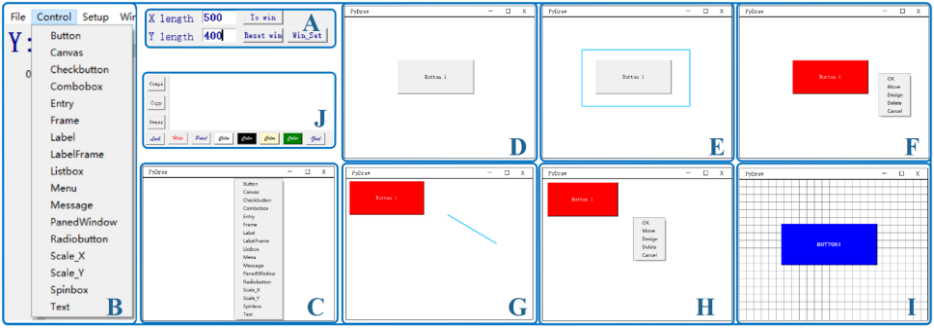

**Fig. 7**   Main steps of drawing GUI by PyDraw

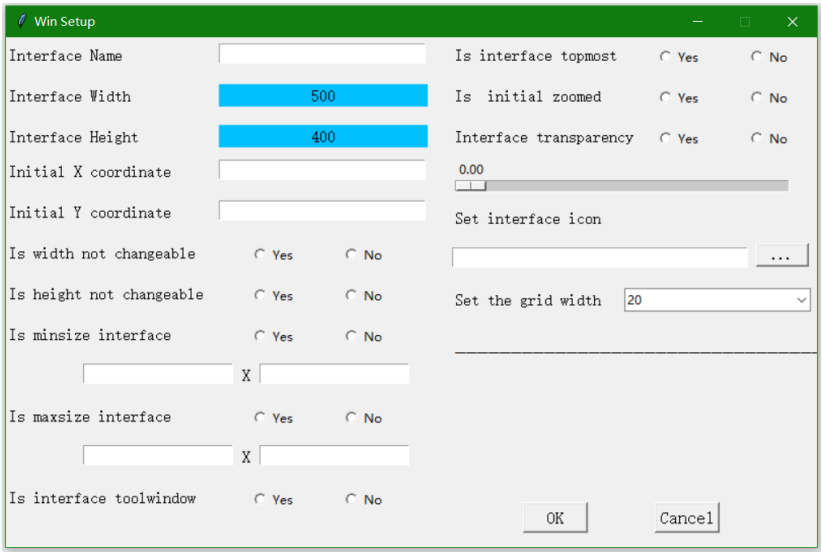

**Fig. 8**   Window Setting panel

***Setting window information:*** As shown in the A sub-graph of Fig. 6, the width and height of the GUI window to be drawn are set in this area. Clicking the "To Win" key to pop up the Window Setting panel as shown in Fig. 7, and then the users can do the detailed window settings in that panel.

***Select the control that want to draw:*** Select the control the user want to draw by clicking on the " Control " key on the main menu or right-clicking on GUI Design Canvas to pop up the control select menu. As shown in Fig. 7, B, C.

***Draw the basic model control:*** Draw the basic model control on GUI Design Canvas by clicking and dragging, as shown in D in Fig. 7.

***Select the control that want to modify:*** Click GUI Design Canvas, drag out the check box, and select the control that want to modify, as shown in E in Fig. 7.

***Selection the function and Execution:*** After releasing the left key, the selected control displays the selected color for markup, GUI Design Canvas pops up the function selection menu, and then the user clicks to select the function to execute, as shown in F in Fig. 7. Move function is selected.

After selecting the "Move" function, click the left mouse button to drag it so that the selected control moves accordingly, and a trajectory line appears to mark the trajectory, as shown in G of Fig. 7. After the move function is completed, the function selection menu reappears for the next operation, as shown in H of Fig. 7. Select "OK" to confirm this operation, select "Cancel" is to cancel this operation. Select "Design" to do the operation of properties and events settings, PyDraw will pop up Converter Text Box, Proties Setting and Events Setting composite panels, users can do the the appropriate properties setting and event configuration for selected control. Throughout the GUI drawing process, the user can click the "Grid" button on the Control Panel at any time to display the grid and assist in GUI drawing design, as shown in J and I of Fig. 7. Users can also click the "Lock" button on the Control Panel at any time to lock GUI Design Canvas so that the GUI Design Canvas can not be moved, to assist the drawing design of the GUI, as shown in J and I of Fig. 7. When a user changes the properties of a control, the control will exist with the changed properties, and the user can still modify the control several times, as shown in I of Fig. 7.

## 3.3 The control layout of PyDraw

To draw a control on GUI Dsign Canvs, we need to set the layout model of the control on GUI Design Canvas. Taking "Button" control as an example, the following is the dictionaris definition form of the basic control "Button" in Tkinter:

*b = Button(master, Various properties definition lists，command=callback)*
*b.pack()*

Among them, the ". *pack ()"* is the layout of the button "b", the control in Tkinter has a variety of layout, and ". *pack ()*" is the most concise. The layout definition code in the GUI Coed generated by "Button" in PyDraw is:

*Button1 = Button(container, Various properties definition lists，command=callback)*
*Button1.place(x=X coordinate, y=Y coordinate)*

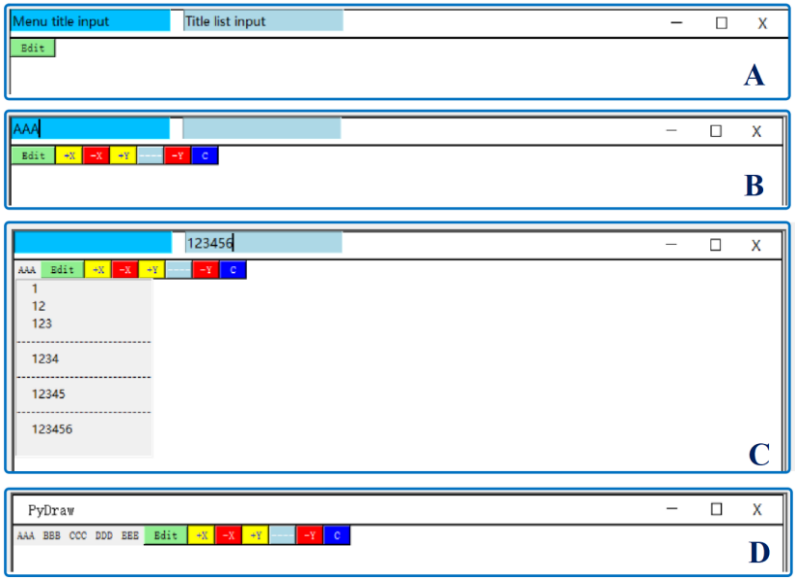

**Fig.9**　Menu edit control

The biggest difference between PyDraw and Tkinter's control definitions is that the current version of PyDraw's control layout is entirely in the form of ". *place (Coordinates definition)* ", which is for the convenience of drawing UI settings. The above "*X coordinates*" and "*Y coordinates*" are the X and Y coordinates of the upper left corner of the control corresponding to GUI Design Canvas when the user draws the control. The future version of PyDraw will add multiple layouts.

### 3.2　Menu edit control

Because tkinter's Canvas control can not directly load the "Menu" control, we designed a Menu editing control to draw the Menu control. After selecting Menu from the control selection menu, the interface shown in A of Fig. 9 appears on GUI Design Canvas. The Menu edit control prompts us to enter the name of the submenu in the first input bar and the name of the list button in the submenu in the second input bar. The editing interface is displayed by clicking "Edit" and recycled by clicking "C", as shown in C and D of Fig. 9. Submenu is added by clicking " +X", submenu is deleted by clicking " -X", list button under corresponding submenu is added by clicking "+Y", list button is deleted by clicking "-Y". Add the dividing line by clicking "-". In PyDraw, "-X, "+Y", "-Y", "----"operations are arbitrary operations that can be performed in disorder, while "+X" operations can only be performed sequentially.

When designing this section, it need to pay attention to the hiding of the submenu list and the position setting when the submenu list is re-displayed. The information to be saved by the processing mechanism is the title of the submenu button and the sequence number in which it appears, as well as the width of the submenu button. Recording the width of the saved submenu button can be used for layout of the submenu list, especially for setting the position coordinates of the upper left corner of the submenu list when the corresponding submenu list is displayed. Among them, the upper left corner coordinates of the corresponding sub menu list of the ith sub menu button are:

$$Xi = \sum_{j=1, j \notin Dict\_D}^{j=i-1} Wj$$

$$Yi = Y0$$

In the above formula: Wj is the X-direction width of the jth submenu button, and Dict_D represents a list of the serial numbers that store the deleted submenus. That is, if the sequence number of the submenu is not in the list, then the submenu still exists, that is, its width needs to be considered. Because the submenu buttons are added in order from small to large, the distance from ith Submenu button to point(0, 0) is the sum of the widths in the X direction of all submenu buttons(form (i-1)th to first) whose ordinal numbers are not added in Dict_D.

The coordinates in the Y direction are the same, that is to say, the menu button forms a row in the X direction. That is, to display a submenu list with a specific sequence number and hide the submenu lists with other sequence numbers, the designer must ingeniously use of the upper left corner coordinates (Xi, Yi) of the submenu list to layout which not in Dict_D. The global dictionary Menu1_Son_Len{} is used to create and save the distance data between the submenu and (0,0) point when processing. Create a global dictionary D_ZhuMenu{} to save the submenu button and the following corresponding submenu list of this submenu and other information, such as the name of the submenu button. Create a global dictionary Menu1_ListCode{} to save the sub menu list to generate GUI Code. The global dictionary Menu1_Delete_Num{} is set up to save the name of the deleted child menu button. Create a global dictionary Menu1{} to save the GUI Code of the corresponding submenu button.

After adding a submenu button or submenu list, or after deleting a submenu button or submenu list, the designer need to update the layout and GUI Code, and store updates of the GUI code. The specific updating algorithm is as follows:

*Add:*
**input:** The currently pressed submenu_button is Button_i, its serial number is i.
The corresponding submenu_list of Button_i is ListBox_i, its serial number is i.
The corresponding submenu_list_button of ListBox_i is BListBox_i_j, its serial number is i_j.

**Update GUI:**
output: Updated Menu layout
if (+X):
    Button(): Create Button_i directly, and use grid() to arrange sequentially in the same line
    Listbox(), Xi: Create new ListBox_i,
        Set the upper left coordinates as X=Xi-1
    D_ZhuMenu{} : Add Button_i and BListBox_i
    Menu1_Son_Len{}, Wi: Add the Wi of Button_i
if (+Y):
    insert() : Add BListBox_i_j into ListBox_i directly

**Update GUI Code:**
output: Updated data of Menu GUI Code, and save it.
if (+X):
    Menu1{} : Save the corresponding submenu GUI Code
if (+Y):
    if j is last:
        Menu1_ListCode{}: Add the GUI Code of BListBox_i_j directly
    else :
        D = {} : Build backup Dictionary
        Backup Menu1_ListCode{} to D{}
        (Total number of D{} is sum, total_number Menu1_ListCode{} is sum+1)
        Copy item j to item sum of D{} corresponding to item j+1 to sum+1
        of Menu1_ListCode{}.

**Fig.10** When added, the algorithm of updating GUI and GUI Code

*Delete :*
**input:** The currently pressed submenu_button is Button_i, its serial number is i.
The corresponding submenu_list of Button_i is ListBox_i, its serial number is i.
The corresponding submenu_list_button of ListBox_i is BListBox_i_j, its serial number is i_j.

**Update GUI:**
output: The GUI Code of Updated Menu layout
if (-X):
    Menu1_Delete_Num.append(i)
    D_ZhuMenu{},destroy(): Button_i and ListBox_i
    del : Item i of D_ZhuMenu{}
if (-Y):
    if j is last:
        D_ZhuMenu{},delete(END): ListBox_i
    else:
        D_ZhuMenu{},delete(j): ListBox_i

**Update GUI Code:**
output: Updated data of Menu GUI Code, and save it.
if (-X):
    pass
if (-Y):
    if j is last:
        del,Menu1_ListCode{}: Delete the GUI Code of BListBox_i_j directly
    else:
        D = {} : Build backup Dictionary
        Backup Menu1_ListCode{} to D{}
        (Total number of D{} is sum, total_number Menu1_ListCode{} is sum-1)
        Copy item j+1 to item sum of D{} corresponding to item j to sum-1
        of Menu1_ListCode{}.

**Fig.11** When deleting, the algorithm of updating GUI and GUI Code

According to the above algorithm, we know that there is no sub-menu button in the GUI Code update when deleting, because PyDraw's Manu edit control only considers sub-menus that are not in Menu1_Delete_Num{} when converting to Menu's GUI Code in order to speed up execution efficiency and system stability. So there's no need for additional GUI Code processing on the deleted submenu buttons, because the GUI Code converter doesn't need to.

### 3.3 Generate GUI code and use it

After drawing the GUI on GUI Design Canvas, the corresponding GUI Code can be displayed in Converter Text Box in real time by clicking the "Compi" key on the Control Panel, and then the user can copy the code directly to use.

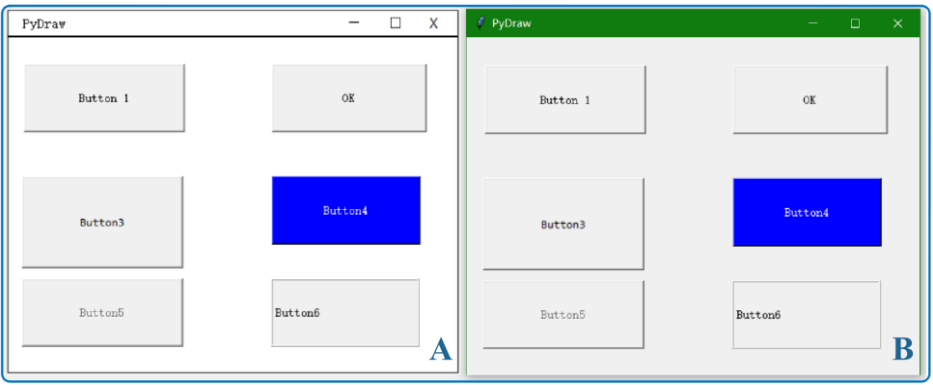

**Fig. 12** Comparison in Button between GUI Design Canvs and actual GUI generation

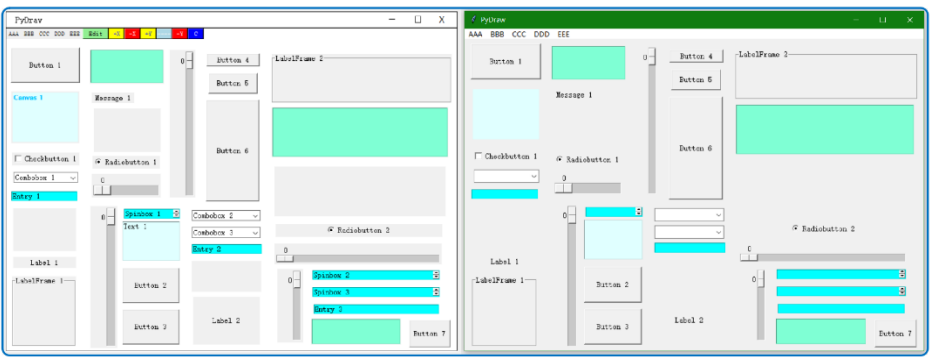

**Fig. 13** Comparison of all kinds of controls in the simulated GUI and the real GUI

**Table 2**  Efficiency comparison between PyDraw generating GUI and writing code generating GUI (unit / sec)

| rows | 2 | 10 | 100 | 200 | 500 | 1000 |
|---|---|---|---|---|---|---|
| *tkinter/s* | 35 | 196 | 1923 | 3991 | 11034 | 25163 |
| *PyDraw/s* | 2 | 16 | 172 | 342 | 892 | 1826 |
| *Tk/PyD* | 17.5 | 12.25 | 11.18023 | 11.66959 | 12.36996 | 13.78039 |

### 3.4 High degree of simulation GUI display

All editing changes in PyDraw are real-time and visualized, and all changes are displayed on GUI Design Canvas in real time. Users can generate GUI Code at any time, and the generated GUI Code corresponds to the GUI design on the GUI Design Canvas at this time. As shown in Fig. 12, the "Button" display of various morphological properties on GUI Design Canvas is highly aligned with the actual GUI display. Among them, A is the simulated GUI and B is the real GUI. PyDraw's GUI Design Canvas uses a white background design to better reflect the original shape of the Tkinter control. Subsequent versions of PyDraw will add multiple window backgrounds. Using the entire Tkinter Canvas control as GUI Design Canvas makes it easier to change the window properties.

PyDraw will give each of the drawn controls sequential names, and users can change the naming of controls. Fig. 13 shows a comparison between the simulated GUI of all PyDraw-enabled controls and the real GUI, where on the left is the analog GUI and the real GUI on the right. Some invisible controls, such as "Frame" and "PanedWindow," are not displayed directly in the actual generated UI, but displayed in PyDraw's GUI Design Canvas. PyDraw can display any drawn controls, which provides great convenience for users to design UI.

### 3.5 Efficiency comparison

To test the efficiency of PyDraw writing GUIs compared with writing code in normal situations, we set up a set of experiments to verify the time spent by PyDraw and writing code manually when generating GUIs with the same number of lines. Experiments show that the time efficiency of GUI design using PyDraw is about 1313% of that written in manual code. As shown in Table 2.

# 4 The programming principles of PyDraw

## 4.1 Kinds of classes

There are 20 classes in PyDraw. The class diagram Fig. 14 has included all the PyDraw classes. The PyDraw (TK) class inherits the Tk window class as the main window class of PyDraw. It includes the main interface settings function Set_UI (self), GUI Code generation and transformation function BianYi (self), and various auxiliary design functions on the Control Panel, as well as the call function when drawing various controls, event configuration functions and so on. The PyDraw class is the main class of PyDraw, and all functions start from here. The SetCK_D class is the aggregate part of the PyDraw (Tk) class, which is used to create the child window. Associated with PyDraw (Tk) is the Hua class, which contains the specific operation functions for drawing various types of controls, as well as the functions Design (self) and UI_Ban_Design (self) for configuration modifications of controls, and the function LuRu_Dict (self) for storing control data in a dictionary. The PyDraw class mainly calls the functions of the Hua class to realize the rendering of GUI. For example, the PyDraw (Tk). BianJI_OK () function calls the Hua. OK () function to validate the function selection menu, and the PyDraw (Tk). Hua_Control () function calls the Hua. Hua_Control () function to draw a specific control, where the Control is the representative of the control type name, such as "Button", "Label" and "Text".

All GUI drawing works on GUI Design Canvas are completed by the corresponding Hua.Hua_Control () function in the Hua class. As the principal class in PyDraw, the implementation of the PyDraw (TK) class relies on the assistance of multiple classes, rather than concentrating all the functions in the PyDraw (TK) class. This facilitates the updating of software functions and software testing and error elimination. For example, we need class Menu_Str to process the GUI Code of Menu menu, class Get_File_Name_GIF to process the'.GIF'file, class Get_File_Name_XBM to process the'.XBM' file, class Chose_Colror to pop up the color selection dialog box and select the color. PyDraw needs Str_ChuLi class to format the string, SJ_Dictionary class to record the event data of the control, Color_Handle class to judge the color of the selected control, XuanDing class to mark and restore the color of the selected control, XuanDing class to process the control after the box selection, SJ_ChuLi class to convert the GUI Code of the control's event. Image_ChuLi class is needed for image name processing and selection and BitMap class is needed for bitmap name processing and selection.

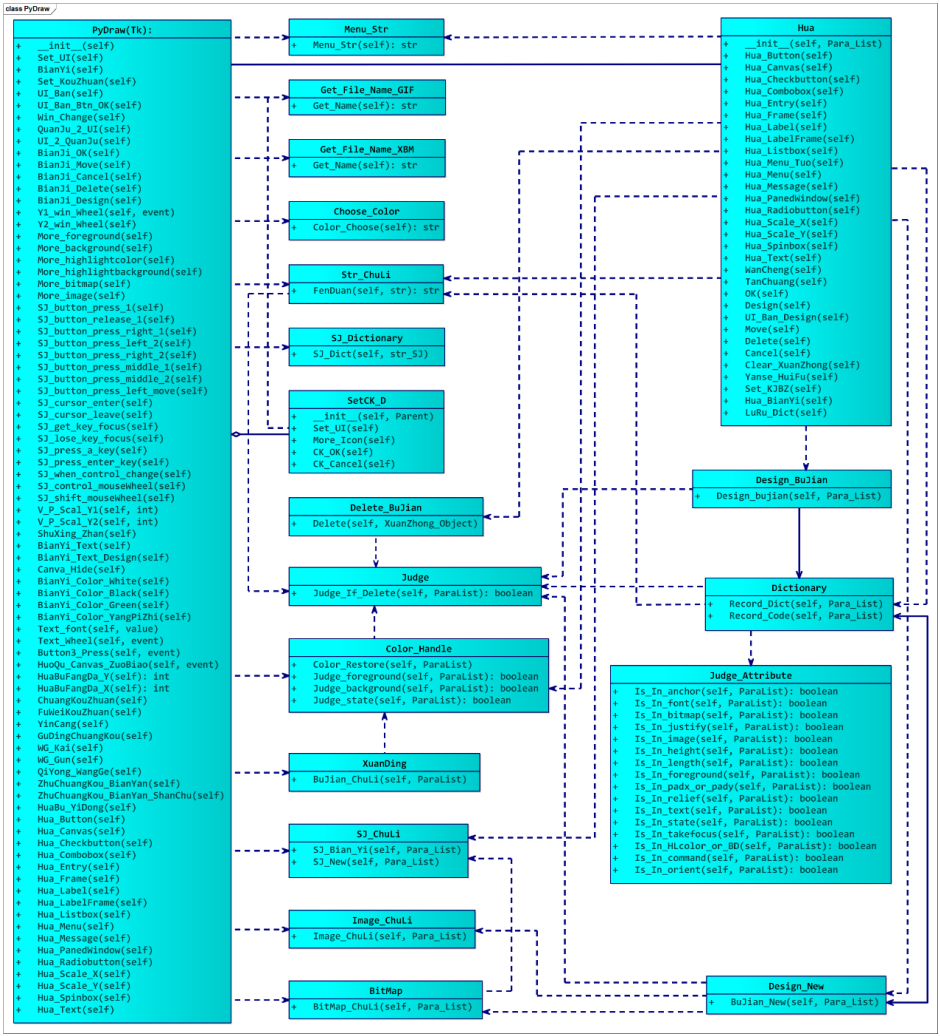

**Fig. 14** The class diagram of PyDraw

As the executor of GUI drawing, Hua class also needs to rely on the assistance of many kinds of classes. For example, Delete class is needed to delete the selected control, Color_Handle class is needed to process the color, SJ_ChuLi class is needed to transform and update the event GUI Code of the control, and Design_BuJian class is needed to draw and design the control. Processing requires the Design_New class to update attributes and events of controls. The Dictionary class saves GUI attribute parameters in a global dictionary, converts the drawn GUI into GUI Code, and enters the dictionary for storage. When saving the property of a control, because the property sets of different types of controls are not necessarily the same, it is necessary to judge whether the control has a certain kind of property value by Judge_Attribute class before entering the property value into the dictionary. Judge classes are primarily used to determine whether a control has been deleted. Judge classes are dependent on many other classes and assist many other classes in performing their functions.

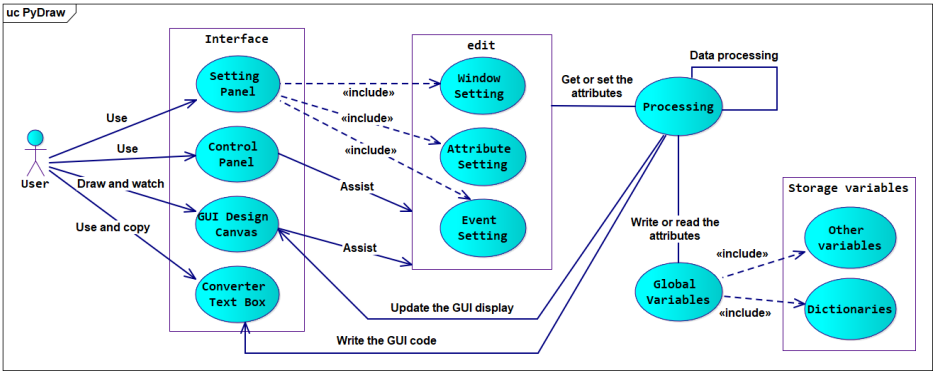

**Fig. 15** The use case diagram of PyDraw

## 4.2 The operation mechanism of PyDraw

We know that PyDraw consists of eight main components: Control Panel, GUI Design Canvas, Converter Text Box, Window Setting, Attribute Setting, Events Setting, Processing, Global Variable. As shown in Fig. 15, Window Setting, Attribute Setting, together with Events Setting constitute Setting Panel. Global Variables includes two categories: Other Variables and Dictionaries. Users can directly manipulate the use of Setting Panel and Control Panel, draw GUI on GUI Design Canvas, view the real-time changes and display of GUI, view the generated GUI Code on Coverter Text Box, and copy the generated GUI Code to use. Control Panel and GUI Design Canvas can also assist users in designing GUI. The GUI editing part of PyDraw consists of three parts: Window Setting, Attribute Setting, and Events Setting. The editing part is connected to the processing mechanism Processing inside the software. All GUI data processing is done on Processing. Processing reads the data from the Global Variables before process the data and then store the processed new data to the Global Variables for updating.

That is to say, PyDraw's data storage is carried out in the form of global variables. Dictionary is the main type of storage variable in global variables because storing data in a dictionary is easy to read and rewrite addressably. This is somewhat similar to the database. Control properties are read using a common set of global variables, of course not all types of controls can read and write to all types of property global variables. Therefore, PyDraw must be able to determine which types of controls have which property variables, which is done through the Judge_Attribute class in Fig. 14. Processing stores the updated data processed in Global Varables and displays the properties changes of the control in real time on GUI Design Canvas. After the user starts GUI Code conversion, Processing converts the GUI drawn at that point into GUI Code and writes it to Converter Text Box in real time.

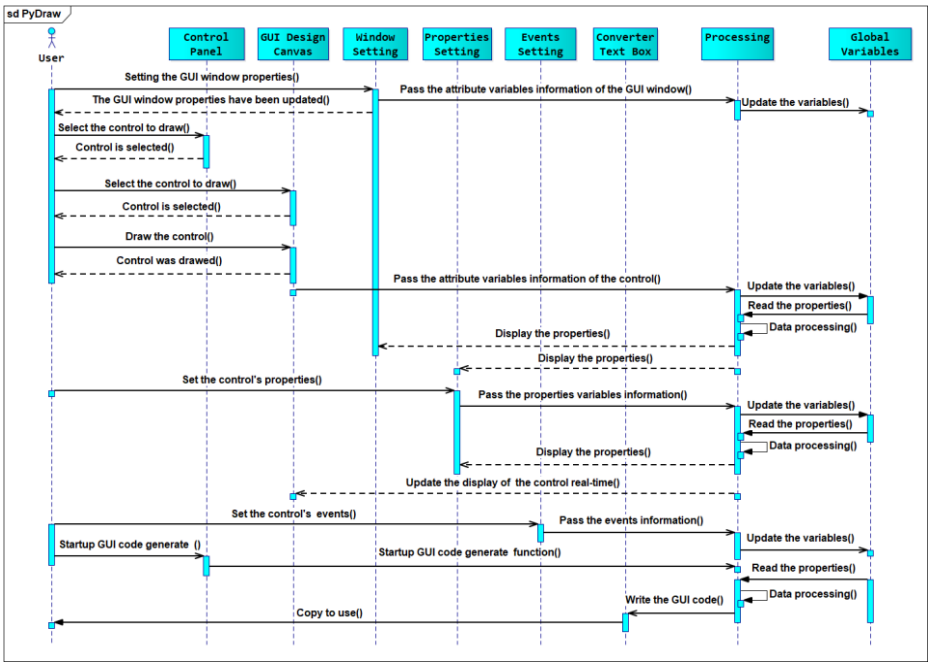

**Fig. 16** The sequence diagram of PyDraw

  Next, we will discuss the specific operation and function realization of PyDraw software system in conjunction with Fig. 16. When drawing GUI with PyDraw, first, the user should sets the parameters of the GUI window through Windows Setting. Then, the software will display the property information of the modified window to tell the user that the property has been modified. At the same time, the parameters set will be saved to the corresponding global variables. Then, the program processing mechanism extracts these parameters and saves them in global variables after processing. The next step is to select the control to be drawn. Users can choose from the Control menu on the Control Panel or from the Control Selection menu that pops up after right-clicking on the GUI Design Canvas. When the user determines the controls to be drawn, it can be drawn on GUI Design Canvs. The information of the drawn control will be passed to the processing mechanism for processing, and then the processing mechanism will save the processed information to the corresponding global variables. At the same time, the corresponding GUI properties information is read in the global dictionary which saves the GUI data information. After processing, the properties of the control are displayed on Windows Setting and Properties Setting. When user updates properties or events of the control, it will be a similar process. The first is to set properties on the Properties Setting panel or control trigger events on the Events Setting panel. When setting properties, after the user sets the new attributes and determines, the updated attributes variable information will be transmitted to the processing mechanism, which will update the corresponding global variables or dictionaries, and read the relevant data of the control, through data processing, and then displayed in the Prooperties Setting. When setting an event, the "Add" button on the Events Setting panel triggers a data transfer event. The control that sets the event and the information about what event it wants to set are passed to the processing mechanism, which updates the event Dictionary of the control immediately. After the GUI drawing design is completed, and even during the whole GUI drawing process, the user can click the "Compi" button on the Control Panel at any time to start the GUI Code generating function, and then the processing mechanism will read the saved GUI data information, process the code data, generate GUI Code, and write the GUI Code to Conveter Text Box,then the user can then copy the GUI Code directly for using.

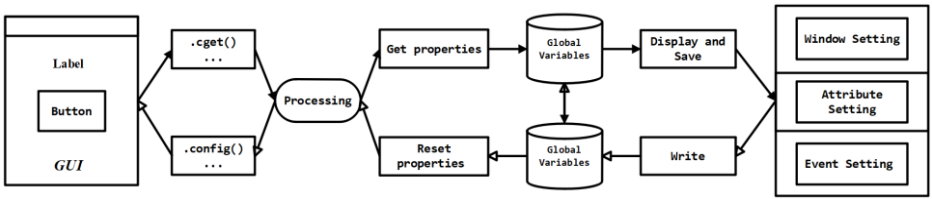

**Fig. 17** The core function of PyDraw and its operation mode.

  From the above analysis, we can see that the user can independently use Windows Setting, Attribute Setting and Events Setting operations at any time from the time the user starts drawing the GUI to the end of drawing the GUI. These three operations are independent and asynchronous. This design also enables users to design GUI more freely and easily.

## 4.3 The core function of PyDraw

**Table 3** Frequently-used properties of Tkinter controls

| | Button | Canvas | Checkbutton | Combobox | Entry | Frame | Label | LabelFrame | Listbox | Message | PanedWindow | Radiobutton | Scale | Text1 | Spinbox1 |
|---|---|---|---|---|---|---|---|---|---|---|---|---|---|---|---|
| container | √ | √ | √ | √ | √ | √ | √ | √ | √ | √ | √ | √ | √ | √ | √ |
| anchor | √ | | √ | | | | √ | | | √ | | √ | | | |
| cursor | √ | √ | √ | √ | √ | √ | √ | √ | √ | √ | √ | √ | √ | √ | √ |
| font | √ | | √ | √ | √ | | √ | √ | √ | √ | | √ | | √ | √ |
| bitmap | √ | | √ | | | | √ | | | | | √ | | | |
| justify | √ | | √ | √ | √ | | √ | | | √ | √ | √ | | | √ |
| image | √ | | √ | | | | √ | | | | | √ | | | |
| width | √ | √ | √ | √ | √ | √ | √ | √ | √ | √ | √ | √ | √ | √ | √ |
| height | √ | √ | √ | | | √ | √ | √ | √ | | √ | | | √ | |
| foreground | √ | | √ | √ | √ | | √ | √ | √ | √ | | √ | √ | √ | √ |
| background | √ | √ | √ | √ | √ | √ | √ | √ | √ | √ | √ | √ | √ | √ | √ |
| padx | √ | | √ | | √ | | √ | √ | | √ | | √ | | √ | |
| pady | √ | | √ | | √ | | √ | √ | | √ | | √ | | √ | |
| relief | √ | √ | √ | | √ | √ | √ | √ | √ | √ | √ | √ | | √ | √ |
| text | √ | | √ | | √ | | √ | √ | | | | √ | | | √ |
| state | √ | √ | √ | √ | √ | | √ | | | | | √ | √ | √ | √ |
| takefocus | √ | √ | √ | √ | √ | √ | √ | √ | √ | √ | | √ | √ | √ | √ |
| highlightcolor | √ | √ | √ | | √ | √ | √ | √ | √ | √ | | √ | √ | √ | √ |
| highlightbackground | √ | √ | √ | | √ | √ | √ | √ | √ | √ | | √ | √ | √ | √ |
| command | √ | | √ | | | | | | | | | √ | √ | | √ |
| length | | | | | | | | | | | | | √ | | |

  The key to drawing a GUI is how to get the parameters of the GUI, how to make update the configuration parameters of the GUI, and how to make the updated GUI display the GUI of the current parameter configuration. Getting the GUI parameters is because we need to master the properties of GUI to understand the current state of GUI. When we understand the status of the GUI, if we need to modify the GUI, then we have to get the GUI parameters to process. After processing of the new parameters, update the GUI parameters and interface to display the refresh. That is to say, there are three core functions to complete the whole process, one is to get the current GUI parameters, one is to update the configuration of GUI parameters, the other is to refresh the current GUI display. The usual GUI frameworks or programming languages combine second functions with third functions. That is to say, refreshing the GUI parameters while refreshing the display of GUI which is simpler and more convenient. Fig. 17 shows that in PyDraw's operating mechanism, there are two core functions, one is the ".cget ()" function, which is used to get the property parameters of the Tkinter control and the other is the ".config ()" function, which updates the property parameters of the configuration Tkinter control and refreshes the display of the control. It can be said that these two functions play a very important role in the Processing of PyDraw. But these two functions alone can't do PyDraw, because PyDraw also needs other GUI parameters in addition to the parameters that these Tkinter controls already have, in order to better achieve GUI rendering and design. For example, x0, y0, name, container four attribute parameters, please refer to the Section 4.4 for details. In addition, there are GUI window parameters, because PyDraw's simulated GUI window is implemented using canvas, so the GUI window parameters need to be set in conjunction with canvas. For the parameters that canvas does not have, additional additions and settings are needed. These are added attribute parameters can not use the ".Cget (") function and the ".Confg ()" function. These parameters will be obtained and updated in the form of global variables in PyDraw, through the customized Processing data processing. As shown in Fig. 17, "..." It represents more implementation methods and functions. In the future, if PyDraw is upgraded and updated, if we get rid of the limitations of the Tkinter framework, then the "..." The functions and methods of delegates will be more colorful.

**4.4 Control properties of PyDraw**

Tkinter supports multiple types of control definitions, but not all kinds of Tkinter controls can have the same number of kinds of property types. Here we have compiled a list of frequently-used properties of Tkinter controls, as shown in Table 3.

As you can see, the Tkinter control has an attribute, length, which is only available to the Scale control. There are also some common attributes that all controls have, which we call global attributes, including width, cursor, and background. In addition to the original Tkinter attributes of the above list, PyDraw adds four attributes for each control: x0, y0, name, container, and some attributes corresponding to the GUI window to facilitate UI design. Add the attribute x0 and Y0 to define the upper left corner coordinates of the control. After experimentation, we find that the coordinates (X, Y) in the layout of the Tkinter control are exactly the upper-left coordinates of the control. Add the name attribute to record and define the name of the control or to modify the name of the control. By default, if the name is not specified, PyDraw will follow something like "Button1", "Button2", "Button3"..., ascending order to names a class of controls automatically. The attribute container is added to record and define the carrier name of the control so as to modify the carrier of the control. The default carrier for PyDraw is "self". Combining Table 3, we know that not all kinds of Tkinter controls can have the same number of kinds of property types. Therefore, we need to define a function to determine which kinds of controls have which kinds of properties in order to better capture and update the properties. In PyDraw, this function is the set of a class of functions. This set of functions is the set of judgement functions in the Judge_Propertyl class in the PyDraw class diagram in Fig. 14.

# 5  PyDraw programming experience summary

(a) *Any GUI generator can be developed with the canvas of any framework or system*:This is in fact consistent with the design philosophy PyDraw wants to convey, through the PyDraw idea, it is easy to develop a GUI and its code generator in a platform with Canvas.

(b) *It is more convenient and quick to use global variables to pass data*: When we design the PyDraw software, the global variables have made a great contribution. Accurate and accurate data transfer and storage of global variables, as well as quick calls between multiple classes and functions, make us to use global variables as the first data transfer type.

(c) *Separate the GUI setup function of the software from other function functions*: The effect is obvious, the designer can effectively separate the management of software GUI display and other software functions to improve efficiency.

(d) *An properties judgment function is needed in the processing mechanism*: To design an efficient GUI editor, the designer must know very well what properties each control has. This allows the designer to config properties, update properties, and GUI code generation. Only the processing mechanism know what properties the control has and how to deel with them,the software will be more efficient. Each GUI generator has multiple controls, many of which have similar settings and properties, so the processing should be centralized. Because attribute types are not necessarily the same, it is very necessary to treat them differently when dealing with them in a centralized way.

(e) *In the processing mechanism, it is desirable to have functions that acquire and set the properties of controls or window, and to refresh the GUI*: To design, modify and update GUI, the designer must understand the status of the current GUI property in real time. Functions that get and set GUI property values make the entire GUI generator design project much easier and faster. For no existing attribute settings and acquisition functions, we need to independently define the corresponding functions, in order to make the GUI generator project proceed smoothly. As mentioned in the 4.3 part, the designer must also have the GUI refresh function. GUI refresh function is also an important function to display human-machine interaction and real-time design.

(f) *A bigger challenge: a purely graphic rendering of PyDraw*: Before we decided to use Tkinter's Canvas control to draw the Tkinter control natively, we tried to use Canvas to draw the control GUI directly, but found it very tedious and could not intuitively reflect the actual GUI situation that was finally generated. But we found that directly drawing controls with Canvas is also very beautiful and elegant, which is very consistent with PyDraw's philosophy, but this requires more work. Only by directly using Canvas to draw GUI can we get rid of the limitations of any GUI framework and truly achieve free GUI rendering. Future PyDraw will focus more on the implementation of drawing GUIs purely with brushes, and realize the GUI programming by dragging lines and drawing specific graphics. Making GUI programming more interesting.

# 6   Conclusion and future work

Based on the definition and design concept of PyDraw, this paper introduces the basic design principle of PyDraw and the valuable design concept behind it. What we want to convey is not just a simple GUI generator, but the concept of software design and software engineering automation behind this simple generator. Human beings are constantly pursuing progress. Software design is also a process of constantly pursuing progress. The emergence and evolution of GUI is a good embodiment. A good software design aids should be simple, easy to use, convenient and fast, can give users a relaxed and pleasant atmosphere of the using of the software, at this point PyDraw achieved. Due to time constraints, PyDraw V1.0 is not powerful enough, and there are unavoidable bugs. And we firmly believe that under the guidance of the excellent PyDraw design concept, the future PyDraw will be even stronger, and we look forward to this day. We've opened source PyDraw on GitHub and hope that more interested and ambitious people will join us to make PyDraw more attractive and handsome. The website address of PyDraw on GitHub is https://github.com/JYLinGitHub/PyDraw.

In the future, we hope that PyDraw can be a typical example of the design concept of drawing GUI, more than just adding Tkinter framework to PyDraw, multi-frame compatibility, excellent control centralized, forming a universal and powerful GUI design mechanism. In future research, we'll focus on how to get PyDraw out of the Tkinter framework and have an autonomous GUI design framework. And, we hope, this framework is evolvable, and anyone is free to develop what they think is a good GUI control. To achieve this goal, we are looking for a general expression of Python GUI design, and striving to design a general editing framework to make Python's GUI design more excellent and powerful. In addition, we believe that graphics modular programming is also an important development direction of future automation software engineering design, the future PyDraw will not only be GUI programming, but also strive to achieve universal software programming. The future goal of PyDraw is to enable users to implement common software designs, including GUI designs, with little code to write.